\documentclass[conference, 10pt]{IEEEtran}
\usepackage[english]{babel}
\usepackage{epsfig}
\usepackage{subfigure}
\usepackage{calc}
\usepackage{amssymb}
\usepackage{amstext}
\usepackage{amsmath}
\usepackage{multicol}
\usepackage{pslatex}
\usepackage{apalike}
\usepackage[small]{caption}
\usepackage{stmaryrd}
\usepackage[standard]{ntheorem}
\usepackage{dsfont} %Pour mathds
\usepackage{tabularx}

\DeclareGraphicsExtensions{.eps}

\begin{document}

\title{A new chaos-based watermarking algorithm}

\author{\IEEEauthorblockN{Christophe Guyeux}
\IEEEauthorblockA{\textit{Computer Science Laboratory (LIFC)}\\\textit{University of Franche-Comt\'e}\\ \textit{Rue Engel-Gros, BP 527, 90016 Belfort Cedex, France}\\
\textit{christophe.guyeux@univ-fcomte.fr}}
\and
\IEEEauthorblockN{Jacques M. Bahi}
\IEEEauthorblockA{\textit{Computer Science Laboratory (LIFC)}\\\textit{University of Franche-Comt\'e}\\ \textit{Rue Engel-Gros, BP 527, 90016 Belfort Cedex, France}\\
\textit{jacques.bahi@univ-fcomte.fr}}
}

\maketitle

\begin{abstract}
This paper introduces a new watermarking algorithm based on discrete chaotic iterations. After defining some
 coefficients deduced from the description of the carrier medium, chaotic discrete iterations are used to mix the
  watermark and to embed it in the carrier medium. It can be proved that this procedure generates topological
 chaos, which ensures that desired properties of a watermarking algorithm are satisfied.

\end{abstract}

\section{INTRODUCTION}

\noindent Information hiding has recently become a major security technology,
especially with the increasing importance and widespread distribution of
digital media through the Internet. It includes several techniques, among which is
digital watermarking. The aim of digital watermarking is to embed a piece of
information into digital documents, like pictures or movies for example. This is for a
large panel of reasons, such as: copyright protection, control utilization,
data description, integrity checking, or content authentication. Digital
watermarking must have essential characteristics including imperceptibility
and robustness against attacks. Many
watermarking schemes have been proposed in recent years, which can be
classified into two categories: spatial domain \cite{Wu2007} and frequency
domain watermarking \cite{Cong2006}, \cite{Dawei2004}. In spatial domain
watermarking, a great number of bits can be embedded without inducing too
clearly visible artifacts, while frequency domain watermarking has been
shown to be quite robust against JPEG compression, filtering, noise
pollution, and so on. More recently, chaotic methods have been proposed to
encrypt the watermark, or embed it into the carrier image for security
reasons.

In this paper, a new watermarking algorithm is given. It is based on the commonly
named chaotic iterations and on the choice of relevant coefficients
deduced from the description of the carrier medium. This new algorithm
consists of two basic stages: a mixture stage and an embedding stage. At 
each of these two stages, the proposed algorithm offers additional steps 
that allow the authentication of relevant information carried by the 
medium or the extraction of the watermark without knowledge about the original image.

This paper is organized as follows: firstly, some basic definitions concerning
chaotic iterations is recalled. Then, the
new chaos-based watermarking algorithm is introduced in Section \ref{sec:chaostic algorithm}. 
Section~\ref{CaseStudy} is constituted by the evaluation of our algorithm: a case study is
 presented, some classical attacks are executed and the results are presented and commented on.
The paper ends by a conclusion section where our contribution is summarized,
and planned future work is discussed.

\section{BASIC RECALLS: CHAOTIC ITERATIONS}

\noindent In the sequel $S^{n}$ denotes the $n^{th}$ term of a sequence $S$, $V_{i}$
denotes the $i^{th}$ component of a vector $V$ and $f^{k}=f\circ ...\circ f$
denotes the $k^{th}$ composition of a function $f$. Finally, the following
notation is used: $\llbracket1;N\rrbracket=\{1,2,\hdots,N\}$.

Let us consider a \emph{system} of a finite number $\mathsf{N}$ of \emph{%
cells}, so that each cell has a boolean \emph{state}. Then a sequence of
length $\mathsf{N}$ of boolean states of the cells corresponds to a
particular \emph{state of the system}. A sequence which elements belong in $%
\llbracket 1;\mathsf{N} \rrbracket $ is called a \emph{strategy}. The set of
all strategies is denoted by $\mathbb{S}.$

\begin{definition}
Let $S\in \mathbb{S}$. The \emph{shift} function is defined by $\sigma
:(S^{n})_{n\in \mathds{N}}\in \mathbb{S}\longrightarrow (S^{n+1})_{n\in %
\mathds{N}}\in \mathbb{S}$ and the \emph{initial function} $i$ is the map
which associates to a sequence, its first term: $i:(S^{n})_{n\in \mathds{N}%
}\in \mathbb{S}\longrightarrow S^{0}\in \llbracket1;\mathsf{N}\rrbracket$.
\end{definition}

\begin{definition}
\label{Def:chaotic iterations}
The set $\mathds{B}$ denoting $\{0,1\}$, let $f:\mathds{B}^{\mathsf{N}%
}\longrightarrow \mathds{B}^{\mathsf{N}}$ be a function and $S\in \mathbb{S}
$ be a strategy. Then, the so-called \emph{chaotic iterations} are defined by $x^0\in \mathds{B}^{\mathsf{N}}$ and $\forall n\in \mathds{N}^{\ast },\forall i\in \llbracket1;\mathsf{N}\rrbracket,$

\begin{equation}
x_i^n=\left\{ 
\begin{array}{ll}
x_i^{n-1} & \text{ if }S^n\neq i \\ 
\left(f(x^{n-1})\right)_{S^n} & \text{ if }S^n=i.%
\end{array}%
\right.%
\label{chaotic iterations}
\end{equation}
\end{definition}

\section{A NEW CHAOS-BASED WATERMARKING ALGORITHM}
\label{sec:chaostic algorithm}

\subsection{Most and Least Significant Coefficients}

\noindent Let us first introduce the definitions of most and least significant
coefficients of an image.

\begin{definition}
\label{def:MSC}
For a given image, the most significant coefficients (in short MSCs), are
coefficients that allow the description of the relevant part of the image, 
\emph{i.e.} its most rich part (in terms of embedding information), through
a sequence of bits.
\end{definition}

For example, in a spatial description of a grayscale image, a definition of
MSCs can be the sequence constituted by the first three bits of each pixel.

\begin{definition}
By least significant coefficients (LSCs), we mean a translation of some
insignificant parts of a medium in a sequence of bits (insignificant can be
understand as: ``which can be altered without sensitive damages'').
\end{definition}

The LSCs are used during the embedding stage: some of the least significant
coefficients of the carrier image will be chaotically chosen and replaced by
the bits of the (possibly mixed) watermark. 

The MSCs are only useful in case of authentication, mixture and embedding
stages will then depend on them. Hence, a coefficient should not be defined
at the same time both as a MSC and a LSC: the LSC can be altered, while the
MSC is needed to extract the watermark (in case of authentication).

\subsection{Stages of the Algorithm}

\noindent Our watermarking scheme consists of two classical stages: the mixture of
the watermark and its embedding into a cover image.

\subsubsection{Watermark mixture}

\noindent For security reasons, the watermark can be mixed before its embedding. A
common way to achieve this stage is to use the bitwise exclusive or (XOR),
for example, between the watermark and a logistic map. In this paper, we will
introduce a mixture scheme based on chaotic iterations. Its chaotic
strategy will be highly sensitive to the MSCs, in case of an authenticated
watermark~\cite{guyeux10}. For the details of this stage see the Paragraph~\ref{watermark
encryption} in Section \ref{CaseStudy}.

\subsubsection{Watermark Embedding}

\noindent This stage can be done either by applying the logical negation of some LSCs, or by replacing them by the bits of the possibly mixed watermark.

To choose the sequence of LSCs to be changed, a number of integers, less
than or equals to the number $N$ of LSCs, corresponding to a chaotic
sequence $\left( U^{k}\right)_{k}$, is generated from the chaotic strategy
used in the mixture stage and possibly the watermark. Thus, the $U^{k}-th
$ least significant coefficient of the carrier image is either switched, or
substituted by the $k^{th}$ bit of the possibly mixed watermark. In case
of authentication, such a procedure leads to a choice of the LSCs which
are highly dependent on the MSCs.

On the one hand, when the switch is chosen, the watermarked image is 
obtained from the original image, whose LSCs $L = \mathds{B}^{\mathsf{N}}$
are replaced by the result  of some chaotic iterations. Here, the iterate
function is the vectorial boolean negation, defined by $f_{0}:  \mathds{B}^{\mathsf{N}}  \longrightarrow \mathds{B}^{\mathsf{N}}, 
(x_{1}, \hdots,x_{\mathsf{N}})  \longmapsto (\overline{x_{1}},\hdots,%
\overline{x_{\mathsf{N}}})$, the initial state is $L$ and strategy is equal to $\left( U^{k}\right)_{k}$. In this case, it is possible to prove that the whole embedding stage satisfies topological chaos
properties~\cite{guyeux10}, but the original medium is needed to extract the watermark.

On the other hand, when the selected LSCs are substituted by the watermark,
its extraction can be done without the original cover. In this case, the
selection of LSCs still remains chaotic, because of the use of a chaotic
map, but the whole process does not satisfy topological chaos~\cite{guyeux10}: the use of
chaotic iterations is reduced to the mixture of the watermark.
See the Paragraph~\ref{Watermark embedding} in Section \ref{CaseStudy} for
more details.

\subsubsection{Extraction}

\noindent The chaotic sequence $U^k$ can be regenerated, even in the case of an
authenticated watermarking: the MSCs have not been changed during the stage
of embedding watermark. Thus, the altered LSCs can be found. So, in case of
substitution, the mixed watermark can be rebuilt and ``decrypted''. In case
of negation, the result of the previous chaotic iterations on the
watermarked image, is the original image.

If the watermarked image is attacked, then the MSCs will change.
Consequently, \emph{in case of authentication }and\emph{\ }due to the high
sensitivity of the embedding sequence, the LSCs designed to receive the
watermark will be completely different. Hence, the result of the decrypting
stage of the extracted bits will have no similarity with the original
watermark.
%
%\subsection{The General Chaos-based Watermarking Algorithm}
%\label{subsec:algo}
%\noindent The different stages of our method can be described in a more general
%framework. First of all, the representation domain of the carrier image has
%to be decided (spatial, DCT, DWT, \emph{etc.}). As explained above, this
%choice will affect the MSCs and LSCs (most and least significant
%coefficients). Then, the question of the mixture of the watermark must be
%asked; if an mixture is needed, then a cipher method will be chosen. 
%For the next stage, we need to know whether the embedding has to be
%authenticated. If so, the significant information contained in the carrier
%image must be summarized in a selection of relevant MSCs; many possibilities
%exist, depending on the representation domain. Then, in case of
%authentication, the MSCs must be associated, on the one hand, to the
%mixture of the watermark and on the other hand, to the embedding of the
%(possibly mixed) watermark. Finally, the LSCs must be defined to
%receive the watermark and the embedding method must be chosen in order to
%alter a few LSCs with the watermark. Each of these stages can be achieved in
%a lot of different ways.
%

\section{A CASE STUDY}
\label{CaseStudy}
\subsection{Stages and Details}

\subsubsection{Images Description}

\begin{figure}[htb]

\begin{minipage}[b]{0.48\linewidth}
  \centering
 \centerline{\includegraphics[width=0.9cm]{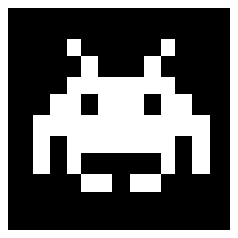}}
  \centerline{{\footnotesize(a) Watermark.}}
\end{minipage}
\begin{minipage}[b]{.48\linewidth}
  \centering
 \centerline{\includegraphics[width=3.6cm]{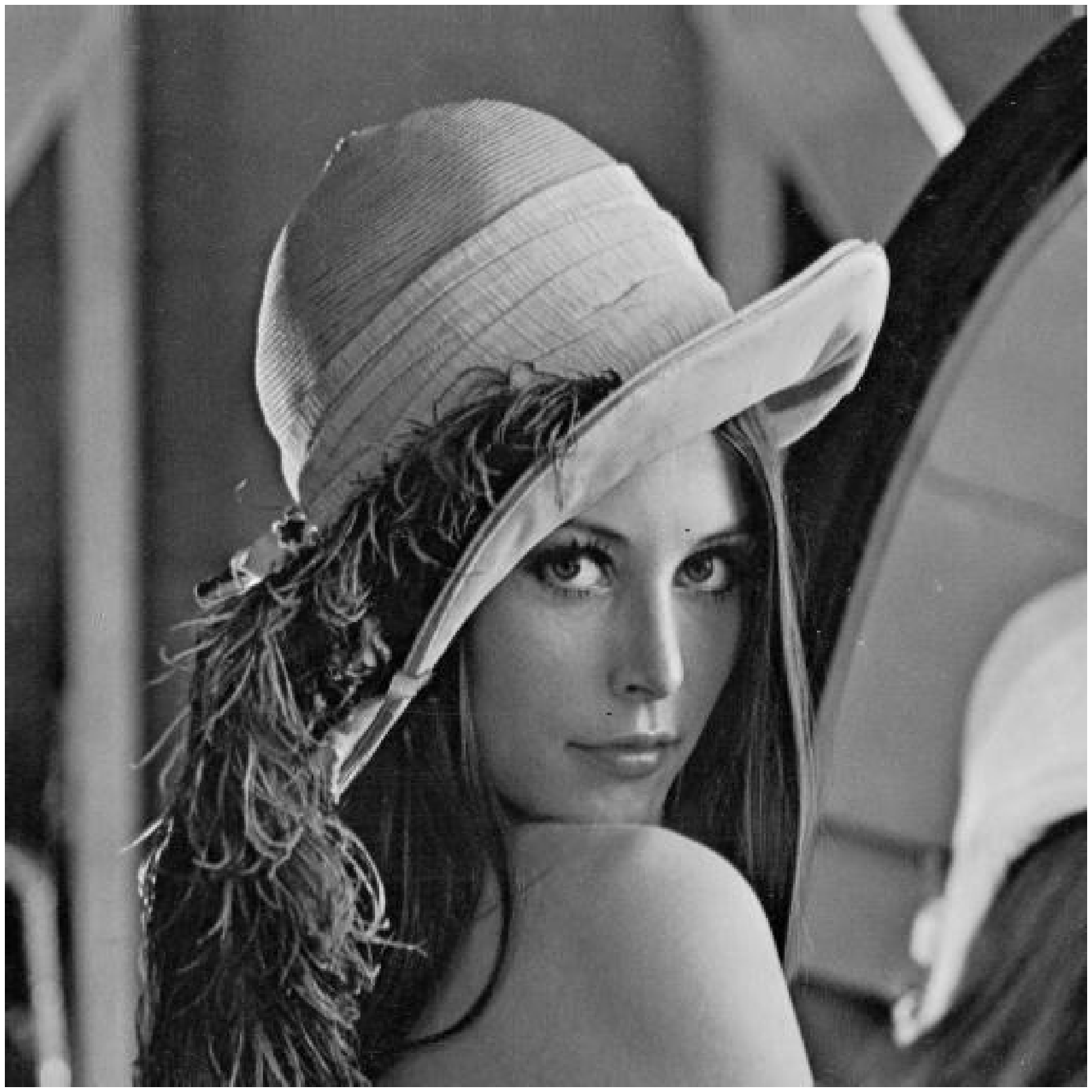}}
  \centerline{{\footnotesize(b) Watermarked Lena.}}
\end{minipage}
\caption{Watermark and watermarked Lena.}
\label{fig:LenaWatermark}
\end{figure}

\noindent Carrier image is the famous Lena, which is a 256 grayscale image and the
watermark is the $64\times 64$ pixels binary image depicted in Fig.~\ref{fig:LenaWatermark}a.
The embedding domain will be the spatial domain. The selected MSCs are the
four most significant bits of each pixel and the LSCs are the three
following bits (a given pixel will at most be modified by four levels of
gray by an iteration). The last bit is then not used.
 Lastly, LSCs of Lena are substituted by the bits of the mixed watermark.

\subsubsection{Mixture of the Watermark}

\label{watermark encryption}

\noindent The initial state $x^{0}$ of the system is constituted by the
watermark, considered as a boolean vector. The iteration function is the
vectorial logical negation $f_{0}$ and the chaotic strategy $(S^{k})_{k\in %
\mathds{N}}$ will depend on whether an authenticated watermarking method is
desired or not, as follows.
A chaotic boolean vector is generated by a number $T$ of iterations of a
logistic map ($(\mu ,U_{0})$ parameters will constitute the private key).
Then, in case of unauthenticated watermarking, the bits of the chaotic
boolean vector are grouped six by six, to obtain a sequence of integers
lower than 64, which will constitute the chaotic strategy. In case of
authentication, the bitwise exclusive or (XOR) is made between the chaotic
boolean vector and the MSCs and the result is converted into a chaotic
strategy by joining its bits as above. Thus, the mixed watermark is the
last boolean vector generated by the chaotic iterations. \newline

\subsubsection{Embedding of the Watermark}

\label{Watermark embedding}

\noindent To embed the watermark, the sequence $(U^{k})_{k\in \mathds{N}}$ of altered
bits taken from the \textsf{M} LSCs must be defined. To do so, the strategy $%
(S^{k})_{k\in \mathds{N}}$ of the mixture stage is used as follows 
\begin{equation}
\left\{ 
\begin{array}{lll}
U^{0} & = & S^{0} \\ 
U^{n+1} & = & S^{n+1}+2\times U^{n}+n ~ (\textrm{mod } \textsf{M}).
\end{array}%
\right.
\end{equation}
To obtain the result depicted in Fig.~\ref{fig:LenaWatermark}b.

Remark that the map $\theta \mapsto 2\theta $ of the torus, which is a
famous example of topological Devaney's chaos \cite{Devaney}, has been chosen
to make $(U^{k})_{k\in \mathds{N}}$ highly sensitive to the chaotic strategy.
As a consequence, $(U^{k})_{k\in \mathds{N}}$ is highly sensitive to the
alteration of the MSCs: in case of authentication, any significant
modification of the watermarked image will lead to a completely different
extracted watermark.

\subsection{Simulation Results}

\noindent To prove the efficiency and the robustness of the proposed algorithm, some
attacks are applied to our chaotic watermarked image. For each attack, a
similarity percentage with the watermark is computed, this percentage is the
number of equal bits between the original and the extracted watermark.

\subsubsection{Zeroing Attack}

\noindent In this kind of attack, some pixels of the image are put to 0. In this case, the results in Table~\ref{Table:Crop} have been obtained. We can conclude that in case of unauthentication, the
watermark still remains after a cropping attack: the desired robustness is
reached. In case of authentication, even a small change of the carrier image leads to a very different extracted watermark.
In this case, any attempt to alter the carrier image will be signaled.

\begin{table}
\centering
{\footnotesize
\begin{tabular}{|c|c||c|c|}
\hline
\multicolumn{2}{|c||}{UNAUTHENTICATION} & \multicolumn{2}{c|}{AUTHENTICATION}
\\ \hline

Size (pixels) & Similarity & Size (pixels) & Similarity \\ \hline

10 & 99.08\% & 10 & 89.81\% \\

50 & 97.31\% & 50 & 54.54\% \\

100 & 92.43\% & 100 & 52.24\% \\ \hline
\end{tabular}
}\\[0pt]
\caption{Zeroing attacks.}
\label{Table:Crop}
\end{table}

\subsubsection{Rotation Attack}

\noindent Let $r_{\theta }$ be the rotation of angle $\theta $ around the center $%
(128, 128)$ of the carrier image. So, the transformation $r_{-\theta }\circ
r_{\theta }$ is applied to the watermarked image.
The good results in Table~\ref{tab:rot} are obtained.

\begin{table}
\centering
{\footnotesize
\begin{tabular}{|c|c||c|c|}
\hline
\multicolumn{2}{|c||}{UNAUTHENTICATION} & \multicolumn{2}{c|}{AUTHENTICATION}
\\ \hline

Angle  & Similarity & Angle & Similarity \\ \hline

5° & 94.67\% & 5° & 59.47\% \\

10° & 91.30\% & 10° & 54.51\% \\

25° & 80.85\% & 25° & 50.21\% \\ \hline
\end{tabular}
}\\[0pt]
\caption{Rotation attacks.}
\label{tab:rot}
\end{table}

\subsubsection{JPEG Compression}

A JPEG compression is applied to the watermarked image, depending on a
compression level. Let us notice that this attack leads to a change of
the representation domain (from spatial to DCT domain). In this case, the
results in Table~\ref{tab:jpeg} have been found.
A good authentication through JPEG attack is obtained. As for the
unauthentication case, the watermark still remains after a compression level
equal to 10. This is a good result if we take into account the fact that we
use spatial embedding.

\begin{table}
\centering
{\footnotesize
\begin{tabular}{|c|c||c|c|}
\hline
\multicolumn{2}{|c||}{UNAUTHENTICATION} & \multicolumn{2}{c|}{AUTHENTICATION}
\\ \hline

Ratio & Similarity & Ratio & Similarity \\ \hline

2 & 82.95\% & 2 & 54.39\% \\

5 & 65.23\% & 5 & 53.46\% \\

10 & 60.22\% & 10 & 50.14\%\\ \hline
\end{tabular}
}\\[0pt]
\caption{JPEG compression attacks.}
\label{tab:jpeg}
\end{table}

\subsubsection{Gaussian Noise}

\noindent Watermarked image can be also attacked by the addition of a Gaussian noise,
depending on a standard deviation. In this case, the results in Table~\ref{tab:gau} have
been found.

\begin{table}
\centering
{\footnotesize
\begin{tabular}{|c|c||c|c|}
\hline
\multicolumn{2}{|c||}{UNAUTHENTICATION} & \multicolumn{2}{c|}{AUTHENTICATION}
\\ \hline

Standard dev. & Similarity & Standard dev. & Similarity \\ \hline

1 & 74.26\% & 1 & 52.05\% \\

2 & 63.33\% & 2 & 50.95\% \\

3 & 57.44\% & 3 & 49.65\% \\ \hline
\end{tabular}
}\\[0pt]
\caption{Gaussian noise attacks.}
\label{tab:gau}
\end{table}

\section{DISCUSSION AND FUTURE WORK}

In this paper, a new way to generate watermarking methods is proposed. The
new procedure depends on a general description of the carrier medium to
watermark, in terms of the significance of some coefficients we called MSC
and LSC. 
Its mixture and also the selection of coefficients to alter are based on
chaotic iterations, which generate topological chaos in the sense of Devaney. Thus, 
the proposed algorithm possesses expected desirable properties for 
a watermarking algorithm. For example, strong authentication of the
carrier image, security, resistance to attacks, and discretion.

The algorithm has been evaluated through attacks and the results expected 
by our study have been experimentally obtained. The aim was not to find the
best watermarking method generated by our general algorithm, but to give a
simple illustrated example to prove its feasibility.
In future work, other choices of iteration functions and chaotic strategies will be explored. They will be compared in order to increase authentication and resistance to attacks. Lastly, frequency domain representations will be used to select the MSCs and LSCs.

\bibliographystyle{apalike}
\bibliography{Secrypt2010}

\end{document}